\documentclass[conference]{IEEEtran}
\IEEEoverridecommandlockouts

\usepackage{cite}
\usepackage{amsmath,amssymb,amsfonts}
\usepackage{algorithmic}
\usepackage{braket}

\usepackage{graphicx} 
\graphicspath{{picture/}}

\usepackage{textcomp}
\usepackage{xcolor}
\usepackage{booktabs}
\usepackage{hyperref}

\def\BibTeX{{\rm B\kern-.05em{\sc i\kern-.025em b}\kern-.08em
    T\kern-.1667em\lower.7ex\hbox{E}\kern-.125emX}}

\begin{document}

\title{Scalable Quantum Molecular Generation \\ via GPU-Accelerated Tensor-Network Simulation}

\author{
\IEEEauthorblockN{
    Yu-Cheng Xiao\IEEEauthorrefmark{2}\IEEEauthorrefmark{3}\IEEEauthorrefmark{1},
    Jen-Yu Chang\IEEEauthorrefmark{2}\IEEEauthorrefmark{4}\IEEEauthorrefmark{1},
    Tzu-Ling Kuo\IEEEauthorrefmark{2}\IEEEauthorrefmark{5},
    Aninda Astuti\IEEEauthorrefmark{2}\IEEEauthorrefmark{6},
    Shu-Chi Wu\IEEEauthorrefmark{9},
    Ka-Lok Ng\IEEEauthorrefmark{6},\\
    Yun-Yuan Wang\IEEEauthorrefmark{8},
    Yu-Ze Chen\IEEEauthorrefmark{3},
    Nan-Yow Chen\IEEEauthorrefmark{2},
    Tai-Yu Li\IEEEauthorrefmark{2}
}

\thanks{\IEEEauthorrefmark{1}These authors contributed equally to this work.}
\IEEEauthorblockA{\IEEEauthorrefmark{2}National Center for High-performance Computing, National Institutes of Applied Research, Hsinchu, Taiwan}
\IEEEauthorblockA{\IEEEauthorrefmark{3}Department of Materials Science and Engineering, National Cheng Kung University, Tainan City, Taiwan}
\IEEEauthorblockA{\IEEEauthorrefmark{4}Department of Electrophysics, National Yang Ming Chiao Tung University, Hsinchu, Taiwan}
\IEEEauthorblockA{\IEEEauthorrefmark{5}Undergraduate Program in Intelligent Computing and Big Data, Chung Yuan Christian University, Taoyuan, Taiwan}
\IEEEauthorblockA{\IEEEauthorrefmark{6}Department of Bioinformatics and Medical Engineering, Asia University, Taichung, Taiwan}
\IEEEauthorblockA{\IEEEauthorrefmark{9}Department of Materials Science and Engineering, National Tsing Hua University, Hsinchu, Taiwan}
\IEEEauthorblockA{\IEEEauthorrefmark{8}NVIDIA AI Technology Center, Taipei, Taiwan}
}
\maketitle

\begin{abstract}
We propose Scalable Quantum Molecular Generation (SQMG), a variational quantum-circuit for sampling molecular graphs using chemical priors on atoms and bonds. SQMG assigns a fixed 3-qubit register to each heavy atom and reuses a single 2-qubit bond register to generate bonds sequentially, yielding an ``atom no-reuse, bond reuse'' architecture with linear qubit scaling. Measurement results are mapped to molecular graphs via lightweight classical decoding with structural constraints. In CUDA-Q, we benchmark the state-vector simulation (CPU/GPU) and the tensor-network simulation (GPU). At $N=8$ heavy atoms, the state-vector simulator (GPU) and the tensor-network simulator (GPU) achieve speeds of up to $4.5\times 10^{4}$ and $2.2\times 10^{3}$ over the state-vector (CPU) baseline, respectively. Crucially, tensor-network simulation extends exact simulation to $N=40$ heavy atoms, where state-vector methods become memory-limited. For training, Bayesian optimization outperforms COBYLA on a Validity$\times$Uniqueness objective, and the same architecture supports \textit{de novo} generation, scaffold decoration, and linker design. Overall, SQMG provides a scalable, reproducible testbed for evaluating accelerated tensor-network simulation and future quantum molecular generation algorithms.
\end{abstract}


\section{Introduction}

Molecular design is central to drug discovery, materials engineering, and chemical synthesis, yet efficiently exploring the vast chemical space remains challenging. In recent years, data-driven generative models such as GANs, VAEs, flow-based approaches, and large language models have learned structural patterns from large molecular databases and produced novel candidates \cite{decao2018molgan,gomez2018automatic,shi2020graphaf,bhattacharya2024large,zdrazil2024chembl,tingle2023zinc}. However, they often require large parameter counts and extensive training data, leading to high computational cost, and may suffer from mode collapse or limited interpretability and controllability \cite{polykovskiy2020molecular,bhattacharya2024large}. This motivates complementary frameworks that prioritize controllability, reproducibility, and scalable evaluation alongside generation quality.

Quantum algorithms offer an alternative route to generative modeling by using parameterized quantum circuits to define nonlinear distributions in an exponentially large Hilbert space and sampling them directly via measurement. On superconducting hardware, microwave control pulses enable high-repetition execution of shallow circuits, enabling high-throughput shot-based sampling \cite{clerk2020hybrid,preskill2018quantum}. Variational quantum algorithms steer the sampling distribution through a quantum--classical loop that estimates an objective from measurements and updates circuit parameters with a classical optimizer \cite{mcclean2016theory}. Building on this paradigm, Chen \textit{et al.} proposed Quantum Molecular Generation (QMG), which incorporates chemical priors into dynamic circuits and applies structural constraints during decoding \cite{chen2025exploring}. However, scaling remains challenging due to dynamic control/reset overhead and the exponential memory and compute costs of state-vector simulation, even with GPU acceleration \cite{gangapuram2024benchmarking,jiao2023communication}.

\begin{figure}[h]
    \centering
    \includegraphics[width=1\linewidth]{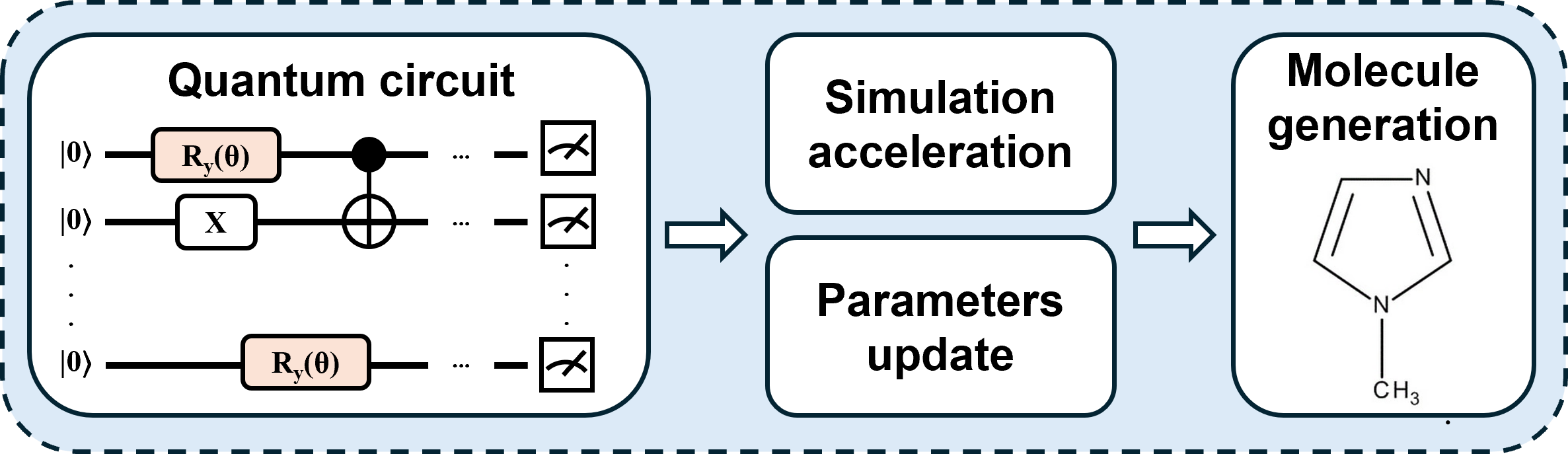}
    \caption{\small Workflow of quantum molecular generation.}
    \label{fig:workflow}
\end{figure}

To address these limitations, we introduce Scalable Quantum Molecular Generation (SQMG), which integrates GPU acceleration with tensor-network (TN) simulation to reduce dynamic overhead and improve scalability. SQMG employs a chemistry-guided variational circuit with an ``atom no-reuse, bond reuse'' architecture, assigning fixed registers to heavy atoms while reusing a bond register to generate bonds sequentially, thereby achieving linear qubit scaling. Implemented in CUDA-Q \cite{nvidia_cudaq,cudaq_backends}, we benchmark state-vector simulation (CPU/GPU) and tensor-network simulation (GPU) \cite{gangapuram2024benchmarking,markov2008simulating,cuquantum2023}. State-vector simulation (GPU) accelerates small-to-medium systems, whereas tensor-network contraction (GPU) alleviates the state-vector memory bottleneck and extends exact simulation to larger qubit counts \cite{markov2008simulating,cuquantum2023}. We further demonstrate \textit{de novo} generation, scaffold decoration, and linker design, positioning SQMG as a reproducible testbed for evaluating GPU/TN-accelerated simulation and future quantum molecular generation algorithms.

\section{Methodology}
\subsection{Architecture for Scalable Quantum Molecular Generation}

\begin{figure*}[t] 
    \centering
    \includegraphics[width=\textwidth]{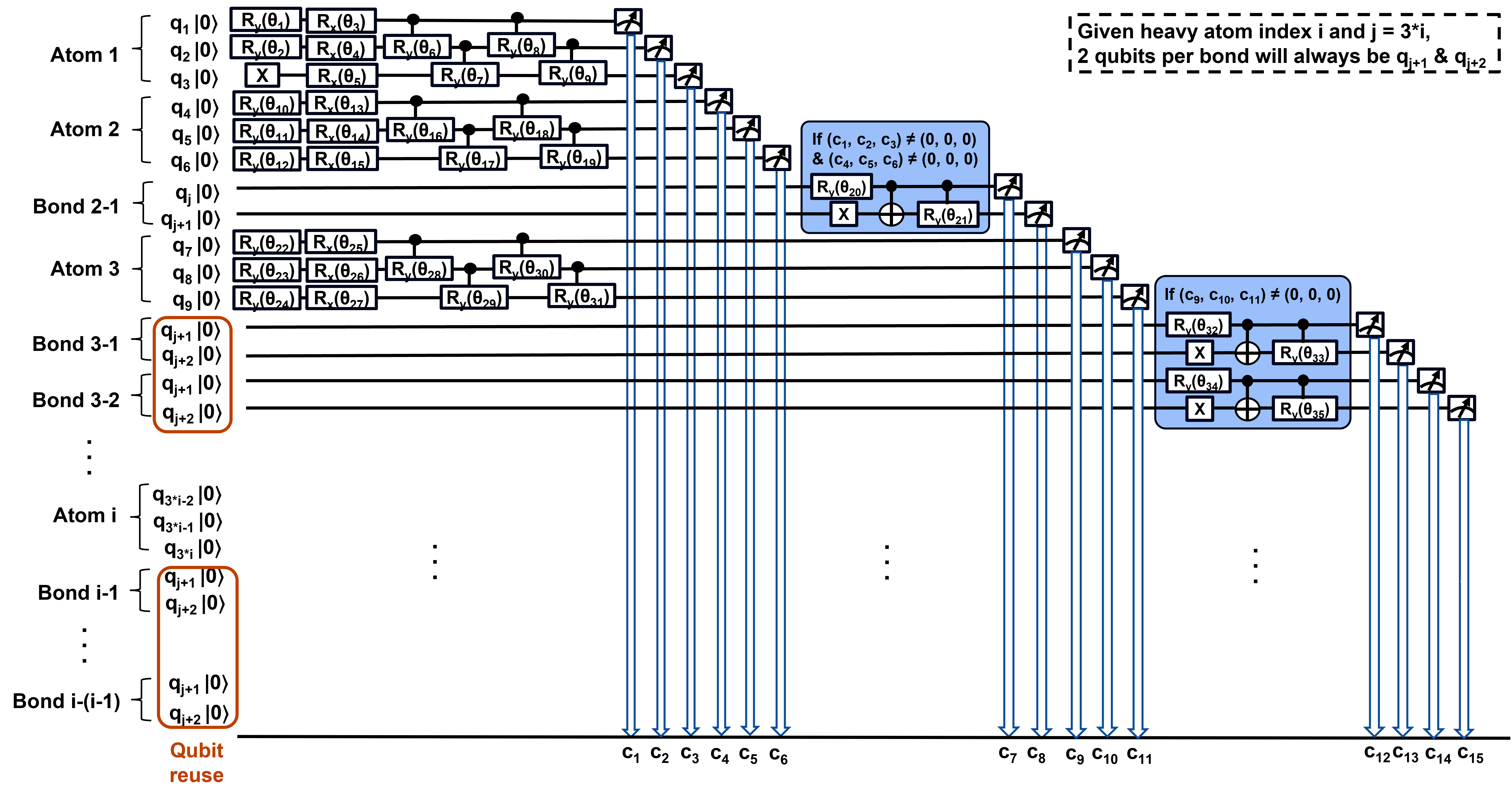} 
    \caption{\small Quantum circuit of the static-atom, dynamic-bond generative ansatz with 3N+2 qubits and bond-qubit reuse.}
    \label{fig:model}
\end{figure*}

The proposed generative ansatz (Fig.~\ref{fig:model}) employs $3i+2$ qubits for a molecule with $i$ heavy atoms, utilizing an \textbf{atom no-reuse, bond reuse} strategy. The circuit is partitioned into two functional subcircuits:

\begin{itemize}
    \item \textbf{Atomic Subcircuit:} Each heavy-atom site is represented by three dedicated qubits, initialized as $\ket{0}$. These undergo chemistry-inspired transformations via parameterized $R_y(\theta)$ rotations and controlled gates. The resulting 3-bit classical codes ($c_1$--$c_6$) map to an 8-state alphabet: one ``NONE'' state and seven heavy-atom types (C, O, N, S, P, F, Cl).  
    \item \textbf{Bond Subcircuit:} If both atoms are present (non-NONE), a conditional bond module \cite{he2019conditional} activates on two shared bond qubits. This generates the bond state ($c_7, c_8$), encoding four possibilities: no bond, single, double, or triple bonds.
\end{itemize}

This design improves efficiency by dynamically reusing bond qubits while keeping atomic identity on static, dedicated qubits \cite{zhang2020qssta}. In contrast, fully dynamic atom-reuse designs can reduce qubit count further \cite{chen2025exploring} but require frequent mid-circuit measurements, classical conditionals, and resets, increasing circuit depth and synchronization overhead and ultimately slowing per-shot sampling.

\begin{table}[h]
    \centering
    \caption{\small Qubit and Parameter Scaling of Static versus Dynamic Circuits as a Function of Heavy-Atom Count}
    \label{tab:performance_comparison}
    \begin{tabular}{@{}cccc@{}}
        \toprule
        \textbf{\begin{tabular}{@{}c@{}} no. of \\ heavy atoms \end{tabular}} & {\textbf{\begin{tabular}{@{}c@{}} no. of qubits\\ required \\ for static circuit \end{tabular}}} & \textbf{\begin{tabular}{@{}c@{}} no. of qubits\\ required \\ for hybrid circuit \end{tabular}} & \textbf{\begin{tabular}{@{}c@{}} no. of \\ parameters \\ in this work \end{tabular}} \\
        \midrule
        2 & 8 & \textbf{8} &  21\\
        3        & 15 &\textbf{11} & 35\\
        4 & 24 & \textbf{14} & 51\\
        5 & 35 & \textbf{17} &  69\\
        10 & 120 & \textbf{32} &  189\\
        20 & 440 & \textbf{62} &  479\\
        30 & 960 & \textbf{92} &  849\\
        40 & 1680 & \textbf{122} &  1959\\
        \( N \)            & \(N(N+2) \) & \( 3N+2 \) & \( N^2+9N-1 \)  \\
        \bottomrule
    \end{tabular}
\end{table}

As summarized in Table~\ref{tab:performance_comparison}, our hybrid ansatz achieves linear qubit scaling compared to the quadratic growth of fully static schemes \cite{gheorghiu2025quantum}. Simultaneously, it maintains a quadratic parameter budget sufficient to capture complex molecular bonding patterns while ensuring a regular, high-throughput sampling architecture.

\subsection{Quantum Circuit Simulation Backends}
To mitigate the exponential cost of state-vector simulation at larger qubit counts, we implement SQMG in CUDA-Q \cite{nvidia_cudaq} and benchmark three backends: \texttt{qpp-cpu} (state-vector, CPU), \texttt{nvidia} (state-vector, GPU), and \texttt{tensornet} (tensor-network, GPU) \cite{markov2008simulating}. State-vector simulation (GPU) provides efficient shot-based sampling for small-to-medium systems, while tensor-network contraction enables scalable exact simulation when state-vector methods become memory-limited.

\subsubsection{State-Vector Simulation (CPU/GPU)}
We use two exact state-vector backends. \texttt{qpp-cpu}, based on Quantum++ \cite{gheorghiu2018quantum}, serves as the CPU baseline for functional verification and small systems, but is limited by the exponential memory and compute cost of storing the full state vector. For higher throughput, we use \texttt{nvidia}, a GPU state-vector backend powered by \texttt{cuStateVec} in the NVIDIA cuQuantum SDK, which accelerates simulation while retaining exact precision \cite{gangapuram2024benchmarking}.

\subsubsection{Tensor-Network Simulation (GPU)}
For larger systems, \texttt{tensornet} simulates circuit evolution via tensor-network contraction \cite{markov2008simulating}, avoiding explicit state-vector materialization. Built on \texttt{cuTensorNet} in the cuQuantum SDK \cite{cuquantum2023}, it enables exact simulation beyond the state-vector memory limit. Tensor-network simulation has also been used to validate large-scale quantum machine learning workloads \cite{chen2024validating}.

\subsection{Optimizers}
\subsubsection{Constrained Optimization BY Linear Approximation (COBYLA)}
To validate trainability and establish an efficient stopping criterion, we optimize the proposed ansatz using COBYLA \cite{virtanen2020scipy}. As a derivative-free method, COBYLA is highly robust to the shot noise inherent in quantum-circuit sampling, circumventing the need for expensive empirical gradient estimation. We directly optimize a composite objective---defined as $\text{Validity} \times \text{Uniqueness}$. Monitoring this optimization trajectory allows us to identify when the model stabilizes. This convergence analysis provides an empirical iteration budget for subsequent experiments, ensuring sufficient training while minimizing unnecessary computational overhead on the quantum backend.

\subsubsection{Bayesian Optimization (BO)}
We employ Bayesian Optimization (BO) \cite{wang2023recent} to tune the continuous parameters of the quantum circuit under a strict evaluation budget. Treating the hybrid quantum-classical pipeline as a gradient-free black box, we construct a Gaussian Process (GP) \cite{deringer2021gaussian} surrogate to model the objective. The GP provides a posterior mean $\mu_t(\mathbf{x})$ and variance $\sigma_t^2(\mathbf{x})$, quantifying both the expected performance and uncertainty at any candidate point $\mathbf{x}$.

To balance the exploration of uncertain regions with the exploitation of high-performing areas, we utilize the Expected Improvement (EI) \cite{zhan2020expected} acquisition function. At each iteration, the GP is updated with the evaluation history $\mathcal{D}_t$, and the next parameter configuration is selected by maximizing EI:
\begin{equation}
    \mathbf{x}_{t+1} = \arg\max_{\mathbf{x}} \mathrm{EI}(\mathbf{x} \mid \mathcal{D}_t).
\end{equation}
This adaptive strategy efficiently identifies optimal circuit configurations while minimizing computationally expensive quantum simulations.

\section{Results}
\subsection{Simulation Time Benchmark}
\begin{figure}[h]
    \centering
    \includegraphics[width=\linewidth]{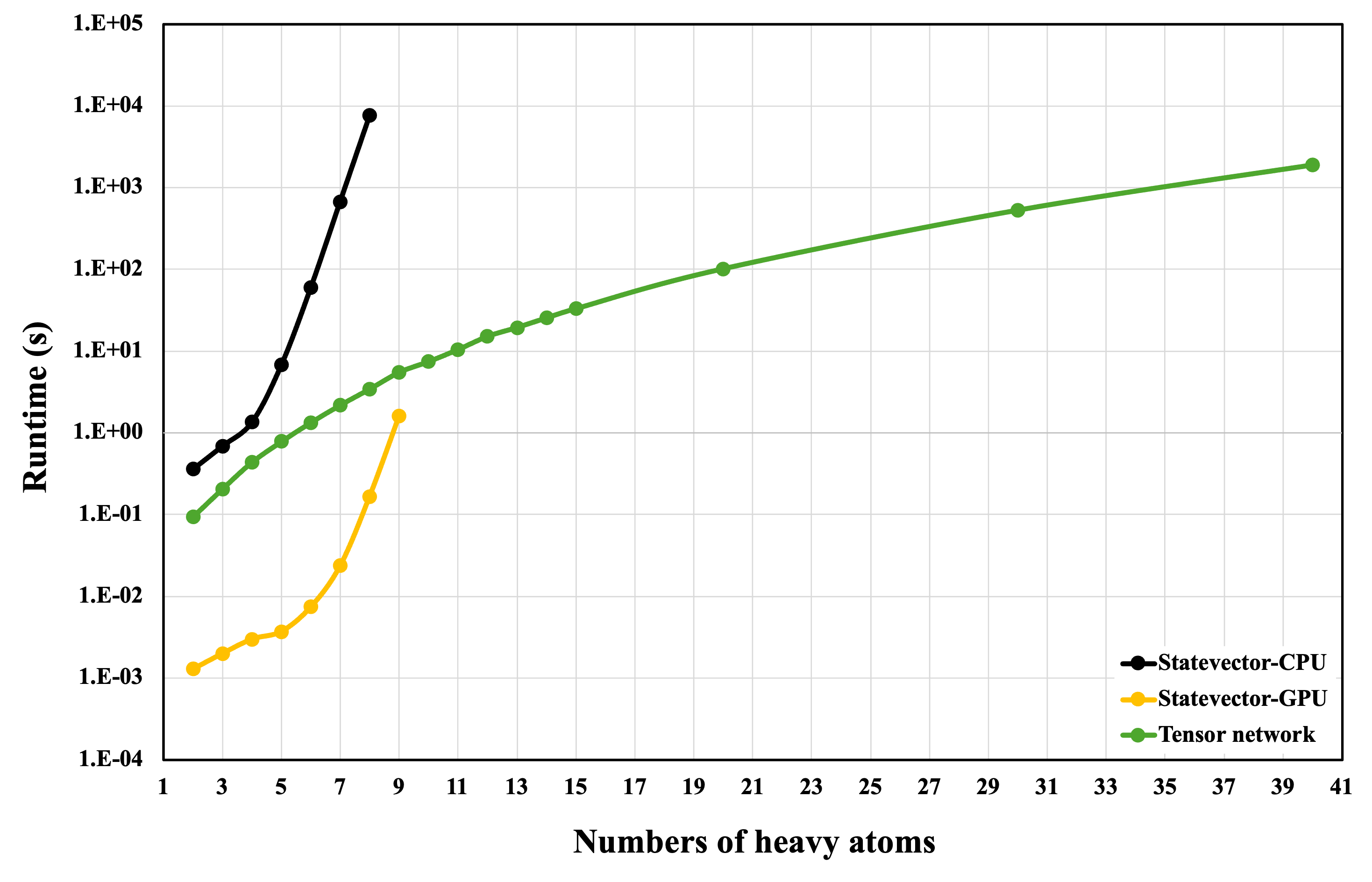}
    \caption{\small Runtime scaling of CUDA-Q simulation backends: state-vector (CPU), state-vector (GPU), and tensor-network (GPU).}
    \label{fig:backend-comparison}
\end{figure}

We observe three distinct scaling regimes as the number of heavy atoms increases (Fig.~\ref{fig:backend-comparison}). All backends are fast for small systems, but they diverge as circuit size grows. State-vector simulation (GPU) eventually becomes memory-limited \cite{jiao2023communication}, state-vector simulation (CPU) exhibits steep exponential growth in runtime, whereas tensor-network simulation (GPU) grows much more gently and remains executable at larger qubit counts.

\begin{enumerate}
    \item \textbf{State-vector simulation (CPU)} scales worst, with steep exponential growth in runtime. While it is fast for very small systems (e.g., $0.01\,\mathrm{s}$ at $N=2$), it becomes impractical quickly, reaching $7{,}639\,\mathrm{s}$ at $N=8$. This behavior follows dense state-vector scaling ($2^{3N+2}$) and is exacerbated by CPU memory-bandwidth and cache limitations.
    
    \item \textbf{State-vector simulation (GPU)} is the fastest option in the small-to-medium regime due to high parallelism and memory bandwidth. At $N=8$, it runs in $0.167\,\mathrm{s}$, outperforming both the CPU baseline and tensor-network simulation. However, it inherits the same $2^{3N+2}$ state-vector memory footprint and becomes infeasible beyond $N=9$ under single-GPU memory constraints.

    \item \textbf{Tensor-network simulation (GPU)} provides the best scalability. Although slower than state-vector simulation (GPU) at small $N$ (e.g., $3.45\,\mathrm{s}$ at $N=8$), its growth is much milder for SQMG circuits by contracting a structured tensor network rather than materializing the full state vector. This enables exact simulations up to $N=40$, making tensor-network simulation the only practical option for SQMG in the large-qubit regime.
\end{enumerate}

At $N=8$ heavy atoms, the runtime differences between backends are pronounced. State-vector simulation (GPU) is approximately $4.5\times 10^{4}$ times faster than the state-vector baseline (CPU), while tensor-network simulation (GPU) is about $2.2\times 10^{3}$ times faster than the CPU baseline.

\begin{figure}[h]
    \centering
    \includegraphics[width=\linewidth]{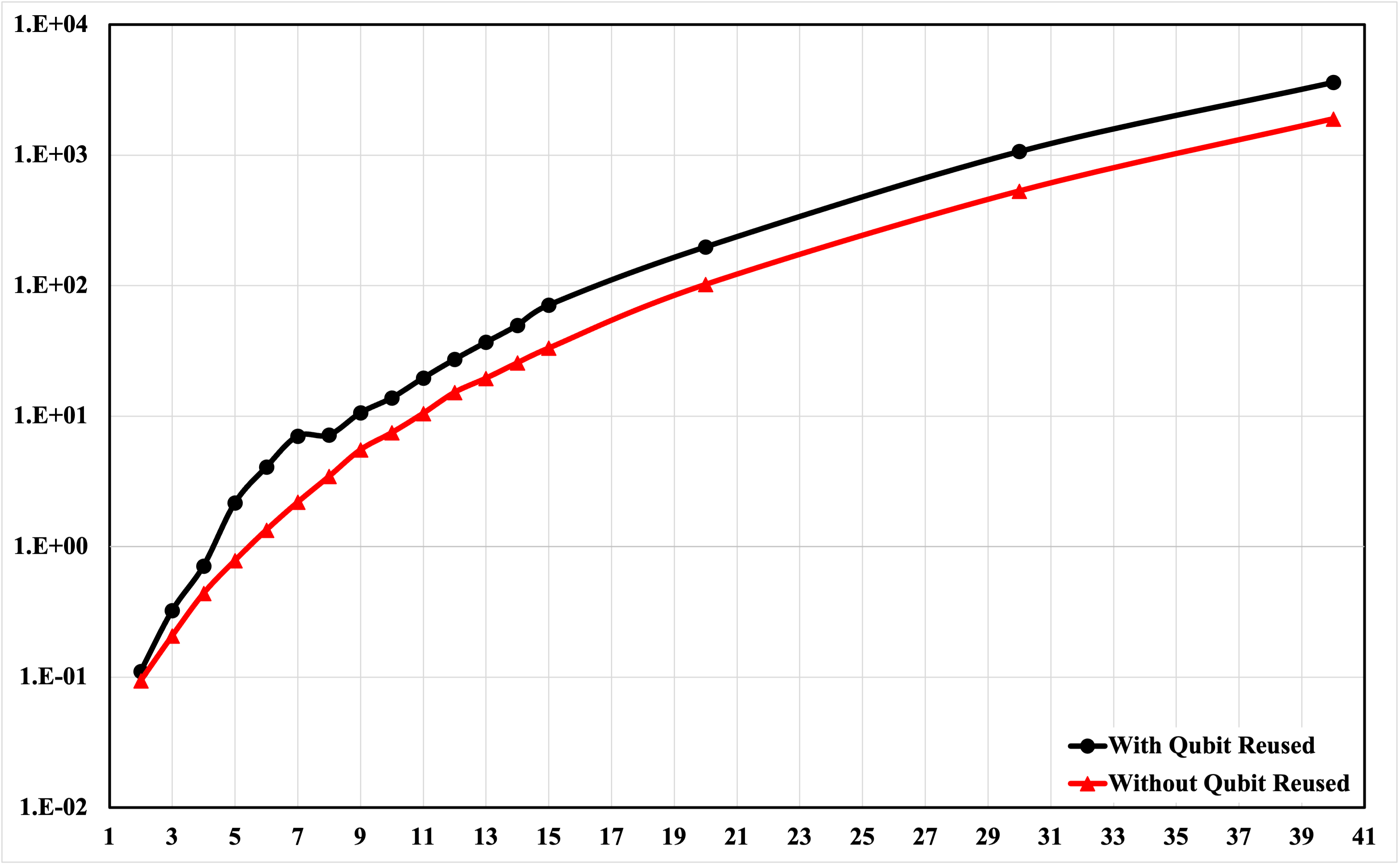}
    \caption{\small Runtime comparison between circuits with and without qubit reuse using the CUDA-Q TensorNet backend.}
    \label{fig:reuse}
\end{figure}

\subsection{Impact of atom with qubit reuse and without qubit reuse}
In this section, we benchmark the proposed quantum circuit under different CUDA-Q backends and under atom no-reuse versus atom reuse configurations, measuring end-to-end execution time as a function of system size.

Fig.~\ref{fig:reuse} presents the runtime scaling of tensor-network simulations under the two configurations. For small systems ($N \leq 6$), the performance difference remains negligible. However, the gap widens significantly as system size increases. At $N=20$, the no-reuse version completes in $102\,\mathrm{s}$, whereas the reuse version requires $198\,\mathrm{s}$. For the largest tested case $N=40$, the reuse configuration reaches $3{,}601\,\mathrm{s}$, nearly doubling the runtime of the non-reuse version $1{,}896\,\mathrm{s}$; in other words, the non-reuse circuit is about $1.9\times$ faster despite using more qubits. This motivates SQMG's atom-no-reuse design: static atomic registers produce more structured circuits that TN backends compress and contract more efficiently.

These results also underscore the advantage of TN simulators (e.g., \texttt{tensornet}), which leverage circuit structure and limited entanglement to scale beyond state-vector memory limits, enabling larger SQMG simulations and more realistic molecular modeling for materials discovery.

\begin{figure}[htbp]
    \centering
    \includegraphics[width=1.0\linewidth]{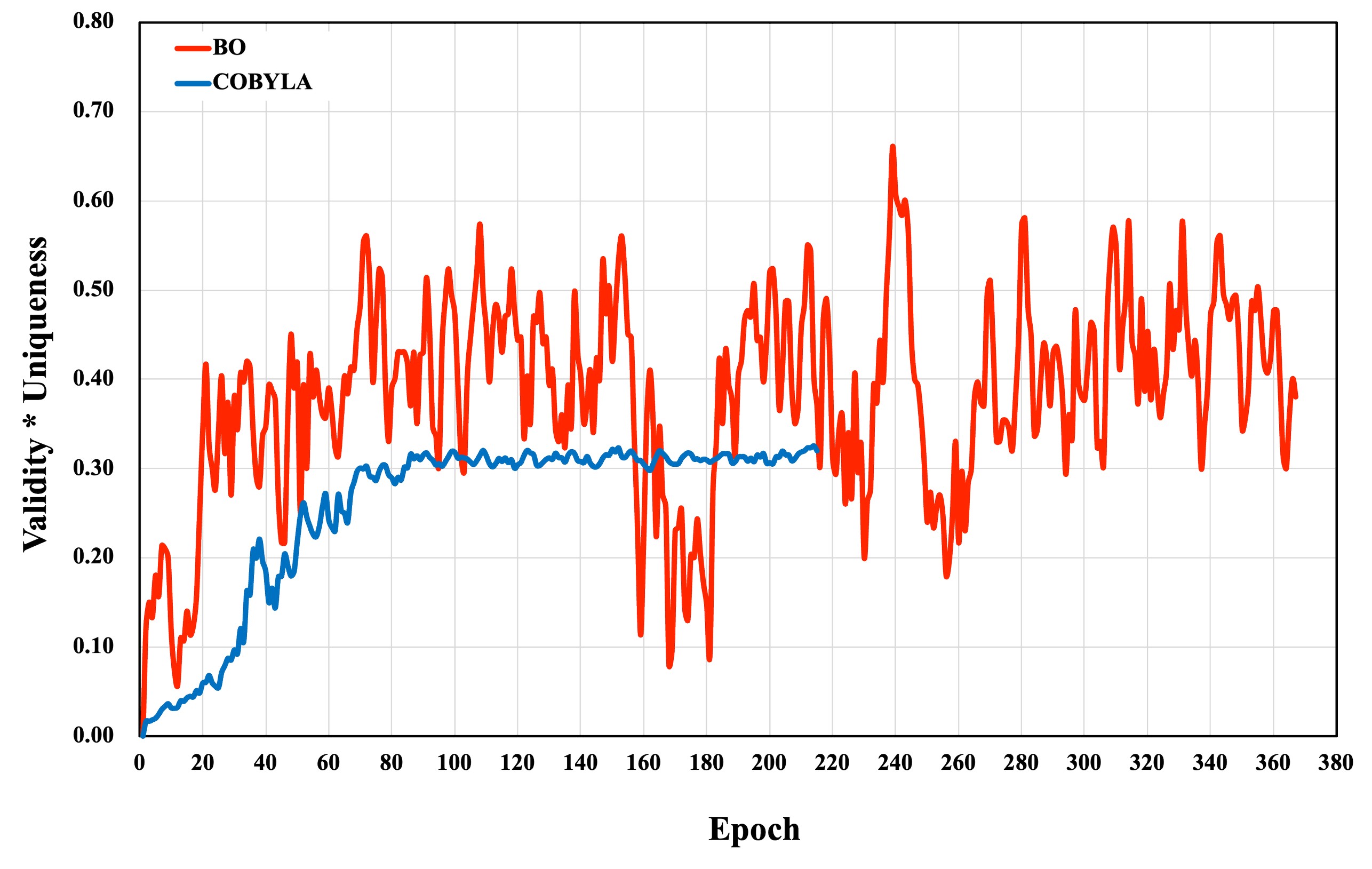}
    \caption{\small Optimization trajectories comparing COBYLA (blue) and Bayesian Optimization (red). While COBYLA converges rapidly to a local optimum, BO exhibits higher variance due to exploration but successfully identifies superior parameter configurations.}
    \label{fig:training}
\end{figure}

\subsection{Optimization Method Comparison}
Fig.~\ref{fig:training} compares the training trajectories of SciPy-based COBYLA and Bayesian optimization (BO) under the composite objective $\text{Validity}\times\text{Uniqueness}$. Curves are reported as a three-epoch moving average to reduce stochastic fluctuations and highlight the optimization trend.

COBYLA improves rapidly at the beginning but saturates around $\text{Epoch}=70$, reaching $0.32$ ($\text{Validity}=0.7100$, $\text{Uniqueness}=0.4507$). This behavior is consistent with COBYLA's local, trust-region search, which can converge once the region contracts around a basin and further local improvements are not found in a noisy, non-convex landscape.

BO exhibits higher variance because its acquisition function balances exploration and exploitation, but its running maximum increases steadily and reaches $0.69$ ($\text{Validity}=0.9600$, $\text{Uniqueness}=0.7188$). Overall, COBYLA provides stable early gains but tends to converge prematurely, whereas BO allocates evaluations more effectively in this noisy, multi-modal setting. We therefore adopt BO as the primary optimizer and retain COBYLA as a local-search baseline.

\begin{figure}[htbp] 
    \centering
    \includegraphics[width=0.5\textwidth]{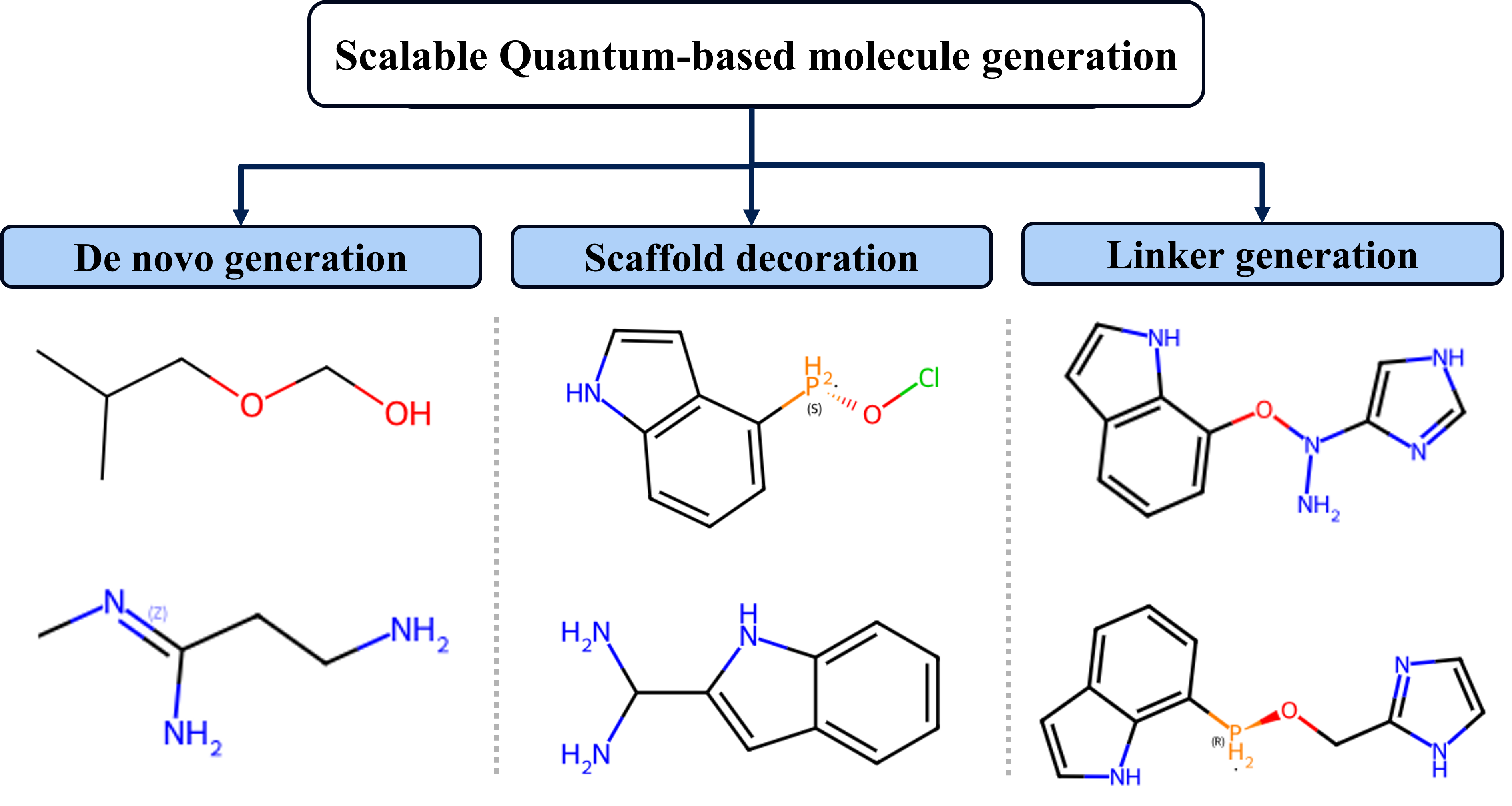}
    \caption{\small De novo generation of full molecules from scratch (left), scaffold decoration of a fixed core with alternative substituents (middle), and linker generation connecting two predefined fragments with quantum-generated linkers (right)}
    \label{fig:model-functionality}
\end{figure}

\subsection{SQMG Functionalities}


SQMG supports three molecular design modes: \textit{de novo} generation, scaffold decoration, and linker generation (Fig.~\ref{fig:model-functionality}). In \textit{de novo} generation, the ansatz samples atom identities and bond patterns across all heavy-atom sites to construct molecular graphs from scratch.

For scaffold decoration, a core scaffold (e.g., indole) is fixed during classical preprocessing by freezing its atoms and bonds, and quantum degrees of freedom are enabled only at predefined substitution sites to select substituents and their connectivity. For linker generation, two terminal fragments are fixed as boundary conditions, and the ansatz acts on intermediate atoms and bonds to generate connecting linkers, exploring linker length, composition, and bond order.

\section{Conclusions}
We presented Scalable Quantum Molecular Generation (SQMG), which combines chemistry-guided circuit design with GPU-accelerated tensor-network (TN) simulation to explore molecular chemical space. Building on prior dynamic QMG-style encodings that use 2 qubits per atom, SQMG assigns a fixed 3-qubit register to each heavy atom, expanding the atomic vocabulary to eight basis states (seven elements plus NONE) while retaining linear qubit scaling at $3N+2$. The circuit follows an ``atom no-reuse, bond reuse'' architecture, storing atomic identity on static registers and reusing a compact 2-qubit bond register to generate bond types sequentially.

We benchmarked state-vector simulation (CPU/GPU) and tensor-network simulation (GPU). At $N=8$ heavy atoms, state-vector simulation (GPU) and tensor-network simulation (GPU) achieved up to $4.5\times 10^{4}$ and $2.2\times 10^{3}$ speedups over the state-vector baseline (CPU), respectively, and tensor-network simulation enabled exact simulation up to $N=40$. Profiling at $N=40$ showed a $\sim 1.9\times$ speedup for the static atomic-register design over fully dynamic atom-reuse variants due to lower TN contraction cost. For training, Bayesian optimization (BO) outperformed COBYLA on the Validity$\times$Uniqueness objective (0.69 vs.\ 0.32).

Finally, SQMG supports \textit{de novo} generation, scaffold decoration, and linker generation. Overall, SQMG provides a scalable and reproducible testbed for evaluating GPU/TN-accelerated simulation backends and future quantum molecular generation algorithms.
\bibliographystyle{ieeetr}
\bibliography{reference} 

@article{decao2018molgan,
  title={MolGAN: An implicit generative model for small molecular graphs},
  author={De Cao, Nicola and Kipf, Thomas},
  journal={arXiv preprint arXiv:1805.11973},
  year={2018}
}

@article{gomez2018automatic,
  title={Automatic chemical design using a data-driven continuous representation of molecules},
  author={G{\'o}mez-Bombarelli, Rafael and Wei, Jennifer N and Duvenaud, David and Hern{\'a}ndez-Lobato, Jos{\'e} Miguel and S{\'a}nchez-Lengeling, Benjam{\'\i}n and Sheberla, Dennis and Aguilera-Iparraguirre, Jorge and Hirzel, Timothy D and Adams, Ryan P and Aspuru-Guzik, Al{\'a}n},
  journal={ACS Central Science},
  volume={4},
  number={2},
  pages={268--276},
  year={2018},
  publisher={ACS Publications}
}

@article{shi2020graphaf,
  title={GraphAF: a flow-based autoregressive model for molecular graph generation},
  author={Shi, Chence and Xu, Minkai and Zhu, Zhaocheng and Zhang, Weinan and Zhang, Ming and Tang, Jian},
  journal={arXiv preprint arXiv:2001.09382},
  year={2020}
}

@article{bhattacharya2024large,
  title={Large Language Models as Molecular Design Engines},
  author={Bhattacharya, Debayan and Cassady, H James and Hickner, Michael A and Reinhart, Wesley F},
  journal={Journal of Chemical Information and Modeling},
  volume={64},
  number={18},
  pages={7086--7096},
  year={2024},
  publisher={ACS Publications}
}

@article{zdrazil2024chembl,
  title={The ChEMBL Database in 2023: a drug discovery platform spanning multiple bioactivity data types and time periods},
  author={Zdrazil, Barbara and Felix, Eloy and Hunter, Fiona and Manners, Emma J and Blackshaw, James and Corbett, Sybilla and de Veij, Marleen and Ioannidis, Harris and Mendez Lopez, David and Mosquera, Juan F and others},
  journal={Nucleic Acids Research},
  volume={52},
  number={D1},
  pages={D1180--D1192},
  year={2024},
  publisher={Oxford University Press}
}

@article{tingle2023zinc,
  title={ZINC 22--A free multi-billion-scale database of tangible compounds for ligand discovery},
  author={Tingle, Benjamin I and Tang, Khanh G and Castanon, Matthew and Gutierrez, John J and Khurelbaatar, Munkhbat and Dandarchuluun, Chinzorig and Moroz, Yurii S and Irwin, John J},
  journal={Journal of Chemical Information and Modeling},
  volume={63},
  number={4},
  pages={1166--1176},
  year={2023},
  publisher={ACS Publications}
}

@article{chen2025exploring,
  title={Exploring Chemical Space with Chemistry-Inspired Dynamic Quantum Circuits in the NISQ Era},
  author={Chen, Lung-Yi and Li, Tai-Yue and Li, Yi-Pei and Chen, Nan-Yow and You, Fengqi},
  journal={Journal of Chemical Theory and Computation},
  volume={21},
  number={13},
  pages={6653--6665},
  year={2025},
  publisher={ACS Publications}
}

@misc{nvidia_cudaq,
  title = {{NVIDIA CUDA-Q}: A comprehensive framework for quantum programming},
  author = {{NVIDIA Corporation}},
  year = {2025},
  howpublished = {\url{https://github.com/NVIDIA/cuda-q}},
  note = {Accessed: 2025-11-30}
}

@misc{cudaq_backends,
  title = {{CUDA-Q} Documentation: Quantum Simulation Backends},
  author = {{NVIDIA Corporation}},
  year = {2025},
  howpublished = {\url{https://nvidia.github.io/cuda-q/using/backends/index.html}},
  note = {Accessed: 2025-11-30}
}

@article{polykovskiy2020molecular,
  title={Molecular sets (MOSES): a benchmarking platform for molecular generation models},
  author={Polykovskiy, Daniil and Zhebrak, Alexander and Sanchez-Lengeling, Benjamin and Golovanov, Sergey and Tatanov, Oktai and Belyaev, Stanislav and Kurbanov, Rauf and Artamonov, Aleksey and Aladinskiy, Vladimir and Veselov, Mark and others},
  journal={Frontiers in Pharmacology},
  volume={11},
  pages={565644},
  year={2020},
  publisher={Frontiers Media SA}
}

@article{preskill2018quantum,
  title={Quantum computing in the NISQ era and beyond},
  author={Preskill, John},
  journal={Quantum},
  volume={2},
  pages={79},
  year={2018},
  publisher={Verein zur F{\"o}rderung des Open Access Publizierens in den Quantenwissenschaften}
}

@article{mcclean2016theory,
  title={The theory of variational hybrid quantum-classical algorithms},
  author={McClean, Jarrod R and Romero, Jonathan and Babbush, Ryan and Aspuru-Guzik, Al{\'a}n},
  journal={New Journal of Physics},
  volume={18},
  number={2},
  pages={023023},
  year={2016},
  publisher={IOP Publishing}
}

@article{markov2008simulating,
  title={Simulating quantum computation by contracting tensor networks},
  author={Markov, Igor L and Shi, Yaoyun},
  journal={SIAM Journal on Computing},
  volume={38},
  number={3},
  pages={963--981},
  year={2008},
  publisher={SIAM}
}

@article{gheorghiu2018quantum,
  title={Quantum++: A modern C++11 quantum computing library},
  author={Gheorghiu, Vlad},
  journal={PLOS ONE},
  volume={13},
  number={12},
  pages={e0208073},
  year={2018},
  publisher={Public Library of Science}
}

@inproceedings{cuquantum2023,
  author    = {Bayraktar, Harun and Charara, Ali and Clark, David and Cohen, Saul and Costa, Timothy and Fang, Yao-Lung L. and Gao, Yang and Guan, Jack and Gunnels, John and Haidar, Azzam and Hehn, Andreas and Hohnerbach, Markus and Jones, Matthew and Lubowe, Tom and Lyakh, Dmitry and Morino, Shinya and Springer, Paul and Stanwyck, Sam and Terentyev, Igor and Varadhan, Satya and Wong, Jonathan and Yamaguchi, Takuma},
  booktitle = {2023 IEEE International Conference on Quantum Computing and Engineering (QCE)}, 
  title     = {{cuQuantum SDK}: A High-Performance Library for Accelerating Quantum Science}, 
  year      = {2023},
  volume    = {1},
  pages     = {1050-1061},
  doi       = {10.1109/QCE57702.2023.00119}
}

@article{virtanen2020scipy,
  title={SciPy 1.0: fundamental algorithms for scientific computing in Python},
  author={Virtanen, Pauli and Gommers, Ralf and Oliphant, Travis E and Haberland, Matt and Reddy, Tyler and Cournapeau, David and Burovski, Evgeni and Peterson, Pearu and Weckesser, Warren and Bright, Jonathan and others},
  journal={Nature methods},
  volume={17},
  number={3},
  pages={261--272},
  year={2020},
  publisher={Nature Publishing Group US New York}
}

@article{clerk2020hybrid,
  title={Hybrid quantum systems with circuit quantum electrodynamics},
  author={Clerk, AA and Lehnert, KW and Bertet, P and Petta, JR and Nakamura, Y},
  journal={Nature Physics},
  volume={16},
  number={3},
  pages={257--267},
  year={2020},
  publisher={Nature Publishing Group UK London}
}

@article{zhang2020qssta,
  title={qSSTA: A statistical static timing analysis tool for superconducting single-flux-quantum circuits},
  author={Zhang, Bo and Li, Mingye and Pedram, Massoud},
  journal={IEEE Transactions on Applied Superconductivity},
  volume={30},
  number={7},
  pages={1--12},
  year={2020},
  publisher={IEEE}
}

@article{gheorghiu2025quantum,
  title={Quantum resource estimation for large scale quantum algorithms},
  author={Gheorghiu, Vlad and Mosca, Michele},
  journal={Future Generation Computer Systems},
  volume={162},
  pages={107480},
  year={2025},
  publisher={Elsevier}
}

@article{chen2024validating,
  title={Validating Large-Scale Quantum Machine Learning: Efficient Simulation of Quantum Support Vector Machines Using Tensor Networks},
  author={Chen, Kuan-Cheng and Li, Tai-Yue and Wang, Yun-Yuan and See, Simon and Wang, Chun-Chieh and Wille, Robert and Chen, Nan-Yow and Yang, An-Cheng and Lin, Chun-Yu},
  journal={Machine Learning: Science and Technology},
  year={2025}
}

@inproceedings{jiao2023communication,
  title={Communication optimizations for state-vector quantum simulator on CPU+ GPU clusters},
  author={Jiao, Chenyang and Zhang, Weihua and Shen, Li},
  booktitle={Proceedings of the 52nd International Conference on Parallel Processing},
  pages={203--212},
  year={2023}
}

@article{he2019conditional,
  title={A conditional generative model based on quantum circuit and classical optimization},
  author={He, Zhimin and Li, Lvzhou and Zheng, Shenggen and Huang, Zhiming and Situ, Haozhen},
  journal={International Journal of Theoretical Physics},
  volume={58},
  number={4},
  pages={1138--1149},
  year={2019},
  publisher={Springer}
}

@article{wang2023recent,
  title={Recent advances in Bayesian optimization},
  author={Wang, Xilu and Jin, Yaochu and Schmitt, Sebastian and Olhofer, Markus},
  journal={ACM Computing Surveys},
  volume={55},
  number={13s},
  pages={1--36},
  year={2023},
  publisher={ACM New York, NY}
}

@article{deringer2021gaussian,
  title={Gaussian process regression for materials and molecules},
  author={Deringer, Volker L and Bart{\'o}k, Albert P and Bernstein, Noam and Wilkins, David M and Ceriotti, Michele and Cs{\'a}nyi, G{\'a}bor},
  journal={Chemical reviews},
  volume={121},
  number={16},
  pages={10073--10141},
  year={2021},
  publisher={ACS Publications}
}

@article{zhan2020expected,
  title={Expected improvement for expensive optimization: a review},
  author={Zhan, Dawei and Xing, Huanlai},
  journal={Journal of Global Optimization},
  volume={78},
  number={3},
  pages={507--544},
  year={2020},
  publisher={Springer}
}

@article{gangapuram2024benchmarking,
  title={Benchmarking quantum computer simulation software packages: State vector simulators},
  author={Gangapuram, Amit Jamadagni and L{\"a}uchli, Andreas and Hempel, Cornelius},
  journal={SciPost Physics Core},
  volume={7},
  number={4},
  pages={075},
  year={2024}
}

\end{document}